\documentclass[prd,nofootinbib,floats,aps,twocolumn,tightenlines,superscriptaddress]{revtex4-1}

\usepackage{graphicx}
\usepackage{bbold}
\usepackage{hyperref}

\usepackage{bm}

\usepackage{amsfonts,amsmath,amssymb,amsthm}

\allowdisplaybreaks

\newcommand{\be}{\begin{equation}}
\newcommand{\ee}{\end{equation}}
\newcommand{\raro}[1]{\mathcal #1}
\newcommand{\ii}{\mathrm{i}}
\renewcommand{\vec}[1]{\bm #1}

\newcommand{\lplanck}{{\ell_\textsc{p}}}

\def\s{\hat{\sigma}}
\def\a{\hat{a}_{\vec{n}}}

\usepackage[usenames,dvipsnames]{xcolor}

\begin{abstract}{We identify a signature of quantum gravitational effects that 
survives from the early universe to the current era: Fluctuations of quantum fields 
as seen by comoving observers are significantly influenced by the history of the 
early universe. In particular we show how the existence (or not) of a quantum 
bounce leaves a trace in the background quantum noise that is not damped and 
would be non-negligible even nowadays. Furthermore, we estimate an upper 
bound for the typical energy and length scales where quantum effects are 
relevant. We discuss how this signature might be observed and therefore used 
to build falsifiability tests of quantum gravity theories. }
\end{abstract}

\begin{document}

\title{Echo of the Quantum Bounce}

\author{Luis J. Garay}
\affiliation{Departamento de F\'isica Te\'orica II, Universidad Complutense de Madrid, 28040 Madrid, Spain}
\affiliation{Instituto de Estructura de la Materia, CSIC, 28006 Madrid, Spain}

\author{Mercedes Mart\'in-Benito}
\affiliation{Perimeter Institute for Theoretical Physics, Waterloo, Ontario N2L 2Y5, Canada}

\author{Eduardo Mart\'in-Mart\'inez}
\affiliation{Perimeter Institute for Theoretical Physics, Waterloo, Ontario N2L 2Y5, Canada}
\affiliation{Institute for Quantum Computing, University of Waterloo, Waterloo, Ontario, N2L 3G1, Canada}
\affiliation{Department of Applied Mathematics, University of Waterloo, Waterloo, Ontario, N2L 3G1, Canada}

\maketitle

\section{Introduction}

In order to test whether a proposal for quantum gravity is correct
or not, it is necessary to derive predictions and contrast them with
observations. Cosmology might be one of the few (if not the only) accessible 
windows to observe  quantum gravity effects, as they 
might be important for the early universe, when matter densities approach Planck scales.
In this paper we explore a way to asses the 
strength of the  signatures of quantum gravity that might be observed nowadays.

 Our proposal can be applied in general to find signatures of any early-universe deviations of 
 standard  General Relativity (GR), i.e. post-Einsteinian corrections to the classical geometry. This said, as a first working scenario we concentrate 
on loop quantum cosmology (LQC) 
 \cite{Bojowald:2008zzb,Banerjee:2011qu,Ashtekar:2011ni}. LQC adapts the 
methods of loop quantum gravity to quantize cosmological 
 systems. Its main result is the replacement of the classical big bang singularity 
 by a quantum bounce \cite{Ashtekar:2006wn}.

In the present study, we analyze the creation of particles measured by a 
particle detector due to the cosmological expansion
when the surrounding matter fields are in the vacuum state.
In cosmology this effect is known as the Gibbons-Hawking effect 
\cite{GibHawking}. Since LQC and GR predict different dynamics for the early 
universe, it is natural to ask whether the probability of excitation today for a 
hypothetical detector switched on since the early universe is different for those 
two dynamics.  In other words, we can ask whether there is any quantum signature  in a   system which has remained coupled to a  
field since the early universe until the current era.  
These effects would have an imprint (via Gibbons-Hawking 
fluctuations and their power spectrum) in the
cosmic microwave background (CMB). See e.g. \cite{Mukhanov:2005sc}.
Naively one would expect that since the two 
dynamics (GR  and LQC) are extremely similar for time scales beyond the Planck 
scale, there is little hope of finding such effects.

Remarkably, we show that these quantum-gravity signatures  can survive until the current era with significant strength, being critically dependent on the specific model of early universe physics. Let us  
consider as the cosmological background a flat, homogeneous, and isotropic 
space-time with compact spatial topology, and a homogeneous massless scalar  
as the matter source. This model is  paradigmatic in LQC since it is solvable 
\cite{Ashtekar:2007em}. Interestingly, one can extract classical effective 
dynamics  coming from this LQC quantum model \cite{Taveras:2008ke}, as it admits dynamical coherent states \cite{Livine:2012mh}.
We will compare the effect of the LQC effective dynamics and that of GR on quantum fields.

With this aim, we introduce a   test matter field 
which, for definiteness we will choose to be a 
conformally coupled massless scalar $\phi$. 
We introduce an Unruh-DeWitt detector \cite{DeWitt} coupled to $\phi$ that will be switched on at some instant of the early universe when both LQC and  GR dynamics are different. 
We shall show that the response of the detector at long times is different for LQC and GR and highly dependent on the early universe physics, and the difference survives regardless of how long we wait, even though both dynamics are distinct only in the brief period when the energy density is of the order of the Planck scale. This will allow us to draw conclusions on the strength and model dependence of the signatures of quantum gravity that we could hope to detect in the current era.

\section{Cosmological background dynamics} We consider a spatially flat, 
homogeneous and isotropic universe (FRW) with  a massless scalar 
$\varphi$ as the matter source. The space-time
geometry $\text{d} s^2=-\text{d}t^2+a^2(t)\text{d} \vec{x}^2$ is  characterized by a scale factor $a(t)$ (we will use  natural units  $c=\hbar=1$).
As usual, for the canonical formalism to be well defined, we need to restrict the 
spatial manifold to a finite region \cite{Ashtekar:2006wn}. For definiteness, we 
will consider a compact three-torus spatial topology, so that the spatial 
coordinates take values in a finite interval $[0,L]$,   the physical volume of the 
universe being $[a(t) L]^3$. This  spatial flat topology is compatible with 
the observational data about our universe \cite{Mukhanov:2005sc}, as long as we consider the scale of 
compactification larger than the observable universe.

The modified Friedmann equation  \cite{Taveras:2008ke}
\begin{align}\label{eff}
\left(\dot a/a\right)^2=
({2\lplanck^2}/{9})\rho 
\left(1- {\rho}/{\rho_\star}\right), \quad \rho_\star:= {\lplanck^2}/({2l^6}).
\end{align}
governs the classical effective dynamics obtained from the LQC model.
Here, $\lplanck=\sqrt{12\pi G}$ is the Planck length;
$\rho=\pi_\varphi^2/(2a^6L^6)$, with $\pi_\varphi$ being the canonical momentum of $\varphi$, is the 
energy 
density of the 
field 
$\varphi$;
 $l$ is a quantization parameter (in LQC the volume has a discrete spectrum  equally spaced by $2l^3$) units\footnote{In the standard LQC literature $l$ is defined as $l:=(4\pi G\gamma \sqrt \Delta)^{1/3}$ \cite{Ashtekar:2009vc}, where $\gamma$ is the Immirzi parameter and $\Delta:=4\pi \gamma \sqrt{3}l_{\text{Pl}}^2$ is the minimum nonzero eigenvalue of the area operator of loop quantum gravity. In this way, $l$ is close to the Planck length $l_{\text{Pl}}:=\sqrt{G\hbar}$.}
\cite{Bojowald:2008zzb,Banerjee:2011qu,Ashtekar:2011ni}; and the critical density 
$\rho_\star$ is the maximum eigenvalue of the density operator in LQC \cite{Ashtekar:2007em}. In the limit $\rho_\star\rightarrow\infty$, 
or equivalently $l\rightarrow 0$, we recover GR.  We will analyze how our results depend on this parameter $l$. $\pi_\varphi$ is a constant of motion whose value is typically chosen to be  $\pi_\varphi\sim10^3$ such that the  dynamics admits an effective semiclassical description throughout the whole evolution \cite{Ashtekar:2006wn}.

\begin{figure}
\begin{center}
\includegraphics[width=0.45\textwidth]{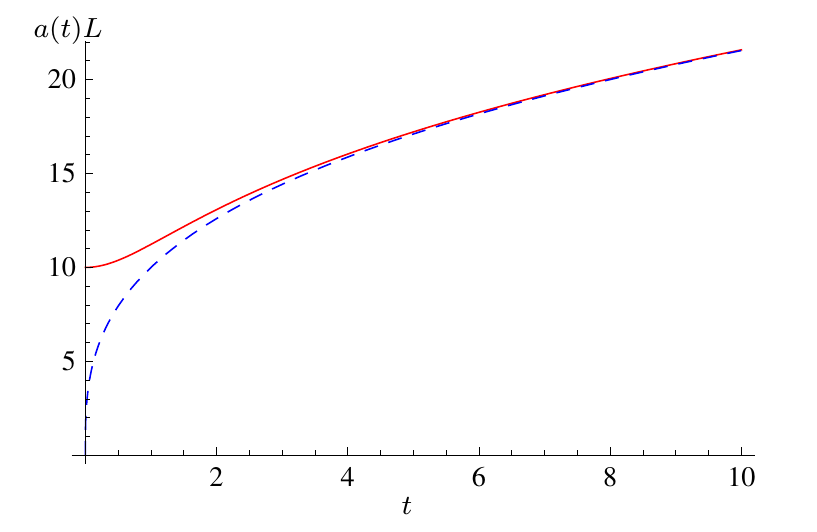}
\caption{Scale factor as a function of the proper time for $l=1$, 
$\pi_\varphi=1000$. The dashed blue curve represents the classical scale factor 
$a_\text{c}(t)L$ and the solid red curve corresponds to the LQC effective scale factor 
$a_\text{q}(t)L$. All quantities are expressed in Planck units, i.e. 
$\lplanck=1$.}
\label{fig:scalefactors}
\end{center}
\end{figure}

The solution to the modified equation \eqref{eff} reads
\begin{align}\label{q}
a_\text{q}(t)=\frac{l}{L}\left(\frac{\pi_\varphi^2}{\lplanck^2}\right)^{1/6}
\left[1+\left(\frac{\lplanck^2}{l^3}t\right)^2\right]^{1/6},
\end{align}
 The classical scale factor, $a_\text{c}(t)=\lim_{l\rightarrow 0} a_\text{q}(t)$, vanishes at 
 $t=0$, and this leads to a big bang singularity. In contrast, the effective solution 
 \eqref{q} never vanishes and takes a positive minimum value 
 $a_\text{q}(0)L=l({\pi_\varphi^2}/{\lplanck^2})^{1/6}$
 (see Fig.~\ref{fig:scalefactors}).
In this case $-\infty < t < \infty$, the universe shrinks for $t<0$, bounces at $t=0$, and expands for $t>0$. The conformal time $\eta_\text{q}$ in terms of the comoving time $t$ is given by an ordinary hypergeometric function:
\begin{align}
\eta_\text{q}(t)=\frac{L}{l}\left(\frac{\lplanck^2}{\pi_\varphi^2}\right)^{1/6} 
\!\!{}_2F_1\left[\frac1{6},\frac1{2},\frac{3}{2},-\left(\frac{\lplanck^2}{l^3}t\right)^2\right]t.
\end{align}
In the limit $t\gg{l^3}/\lplanck^2$, this function behaves as the classical one 
plus a constant, $\eta_\text{q}(t)\rightarrow \eta_\text{c}(t)+\beta$, where
\begin{align}\label{etaclass}
\eta_\text{c}(t)=\frac{3L\,t^{2/3}}{2(\lplanck^2\pi_\varphi^2)^{1/6}},
\quad
\beta=\frac{l^2L\sqrt{\pi}\,\Gamma\left(-\frac1{3}\right)}{(\pi_\varphi^2 
\lplanck^{10})^{1/6}2\Gamma\left(\frac1{6}\right)}.
\end{align}

Effective and classical dynamics are quite 
different at early times, while they agree at late times. A natural question arises: Can 
an observer distinguish---at late times---between both dynamics without having access to energy resources in the Planck scale? 
We will 
answer this question by analyzing the response of a low-energy detector coupled to a test field 
immersed in the above space-time.

\section{Particle production in cosmology} We consider a conformal massless scalar field $\phi$ filling the volume of the universe. We Fock quantize this field 
choosing as vacuum the 
one associated with the following
 complete set of orthonormal modes for the conformally invariant 4-dimensional Klein-Gordon equation
\cite{cosmoq} in 
our cosmological background:
\begin{align}\label{modes}
u_{\vec{n}}(\vec{x},\eta)&= \frac{e^{-\ii\omega_{\vec{n}} \eta}}{a(\eta)\sqrt{2 L^3 \omega_{\vec{n}}}}e^{\ii \vec{k_{\vec{n}}} \cdot \vec{x} }.
\end{align}
Here, \mbox{$\vec{k_{\vec{n}}}= \frac{2\pi\vec{n}}{L}$},  $\omega_{\vec{n}}=\left|{\vec k_{\vec{n}}}\right|$, and \mbox{$\vec{n}=(n_x,n_y,n_z)\in\mathbb{Z}^3$.}
This  vacuum  remains  invariant  as the universe expands, and is usually referred to as ``conformal vacuum'', even though the field has a zero mode (see below).
The field operator is $\hat{\phi} =\sum_{\vec{n}} (\a u_{\vec{n}} +\a^\dagger u^*_{\vec{n}}) $
with $\a$ ($\a^\dagger$) the corresponding annihilation (creation) operators. This Fock  quantization  is the only  one  (up to unitary equivalence) with a vacuum invariant under the spatial 
isometries and unitary quantum dynamics \cite{Gomar:2012xn,Gomar:2012pp}. 
 In the above expression \eqref{modes}, we note that $\bm n\neq \bm 0$. As  discussed in \cite{Gomar:2012xn,Gomar:2012pp},  in the following we will ignore the zero mode, and therefore 
in our calculations we assume that our detector is not coupled to this mode, restricting the sums to $\bm n\neq \bm 0$. The unitary evolution and the uniqueness of the representation do not depend on the removal of a finite number of degrees of freedom. This mode can always be quantized separately. Furthermore, one can see that the coupling of the detector to the zero mode $\vec n=\bm 0$ of the field is not relevant to the effects reported in this article. Something that can be readily seen by, for instance, introducing a small field mass or an IR regularization, or  by coupling the detector to the derivative of the field instead of the field itself \cite{IRdet,Benitoooooo} (a model which is free from IR divergencies).

In a homogeneous and isotropic universe there exists a family of privileged observers, called comoving observers, that see an isotropic expansion from their proper reference frame.  Their proper time  does not coincide with the conformal time. This means that comoving detectors actually detect particles even in the conformal vacuum defined above. This is the well-known Gibbons-Hawking effect \cite{GibHawking}.

A particle detector can be understood as a convenient witness of quantum 
fluctuations \cite{GibHawking,cosmoq}. We will use the Unruh-DeWitt model \cite{DeWitt}, which describes the 
local monopole interaction of a two-level quantum system with a scalar field \cite{Birrell,cosmoq,Reznik2005,AasenPRL}. Although simple, this model encompasses all the fundamental features of the radiation-matter interaction   \cite{Wavepackets}.
The Hamiltonian of the coupled  system  in the interaction picture is $\hat{H}_I(t)=\lambda\;\chi(t)(\s^+ e^{i\Omega t}+\s^- e^{-i\Omega t})\hat{\phi}[\vec{x}(t),\eta(t)]$. Here
$\lambda$ is the coupling strength, $\chi(t)$ is the detector's switching function, $[\vec{x}(t),\eta(t)]$ is the world line of the detector, $\Omega$ is the energy gap between its ground and  excited states, and $\s^\pm$ are $SU(2)$ ladder operators. 

We will consider several switching functions: namely sudden $\chi_1(t)$, linear ramping up $\chi_2(t)$, analytical ramping up $\chi_3(t)$, and analytically smooth activation $\chi_4(t)$, all of them   compactly  supported in the interval $t\in[T_0,T]$:  
\begin{align}
\nonumber \chi_1(t)&=1, \qquad \\
\nonumber\chi_2(t)&=\left\{\begin{array}{ll}
{\min\left[ ({t-T_0})/{\delta},1\right]}& \qquad   t<(T+T_0)/{2} \\
{\min\left[({T-t})/{\delta},1\right]}& \qquad t\geq(T+T_0)/{2},
\end{array}\right.\\
 \chi_3(t)&= \tanh \left(\textstyle{\frac{t-T_0}{\delta}}\right)-\tanh \left(\textstyle{\frac{t-T}{\delta}}\right)+\tanh \left(\textstyle{\frac{T_0-T}{\delta}}\right), \\
\nonumber\chi_4(t) &= \left\{ \begin{array}{ll}
S\left[({t-T_0})/{\delta} \right] \qquad&    t<T_0 +\pi \delta\\
1&  t\in[T_0   +\pi \delta, T - \pi \delta) \\
S\left[ ({T-t})/{\delta}\right] &   t\ge T-\pi \delta ,
\end{array}\right.
\end{align}
where $S(x)=[1-\tanh(\cot x)]/2$ and $\delta$ controls the ramping up.
Although for simplicity we consider a sudden switching $\chi_1(t)$  in several of our calculations, it is known that in $3+1$D this leads to UV divergent integrals which  only depend on the switching and not on the state of the field or the background geometry \cite{Louko:2007mu}. Nevertheless, as we will compute differences in probabilities of detectors with the same switching functions, our results will be devoid of any such switching effect  (including UV divergences) and, furthermore, we show that our results are largely independent of the particular switching function.

We consider the test field to be in the conformal vacuum state defined above and the detector initially in its ground state. The detector is stationary in the comoving frame, $\vec{x}(t)=\vec x_0=(x_0,y_0,z_0)$. Provided that $\lambda$ is small enough, we can compute probabilities perturbatively. At leading order, the probability of transition for the detector, switched on at $T_0$, to be excited at time $T$ is
\begin{align}
\label{prob}
P_\text{e}(T_0,T)& =\lambda^2 {\sum_{\vec{n}}}'|I_{\vec{n}}(T_0,T)|^2 +\mathcal{O}(\lambda^4),\\
\label{none}
I_{\vec{n}}(T_0,T)&=\int_{T_0}^T \text{d}t 
\frac{e^{-\frac{ 2\pi\ii 
\bm n\cdot \bm x_0}{L} } }{a(t)\sqrt{2\omega_{\vec n}L^3}}e^{\ii[\Omega t + \omega_{\vec{n}} 
\eta(t)]},\end{align}
where the prime means that the $\vec n=\vec 0$ mode is excluded from the sum.

\section{Excitation probability of the detector under both space-time dynamics}

 We will compare the probability $P^{\text{q}}_\text{e}(T_0,T)$ of the detector to get excited 
when the universe evolves under the effective LQC dynamics, with the probability  
$P^{\text{c}}_\text{e}(T_0,T)$ of the detector to get excited when the universe evolves 
under the GR dynamics. In particular we will check whether the signatures of the 
behavior at early times  survive or not in the long time regime. 

In order to alleviate the numerical computations, it is convenient to 
split the integrals in \eqref{none} into two intervals: $t\in [T_0,T_\text{m}]$ and 
$t\in [T_\text{m},T]$. $T_\text{m}$ is a short time sufficiently large for
$\eta_\text{q}(T_\text{m})\approx\eta_\text{c}(T_\text{m})+\beta$ as shown in 
\eqref{etaclass}. In other words, we split the integrals in a regime 
where LQC and GR  appreciably predict different dynamics and a regime of long 
times where the dynamics are essentially the same. $T_{\text{m}}$ would 
typically be of the order of few times $l^3/\lplanck^2$, as shown in 
Fig. 
\ref{fig:scalefactors}. Then, the difference between the probabilities, 
$\Delta P_\text{e}(T_0,T)\equiv P^{\text{q}}_\text{e}(T_0,T)-P^{\text{c}}_\text{e}(T_0,T)$, at leading order can be written as
\begin{align}
&\Delta P_\text{e}(T_0,T)=\lambda^2{\sum_{\bm n}}'\bigg[\left|I_{\bm 
n}^\text{q}(T_0,T_\text{m})\right|^2-\left|I_{\bm n}^\text{c}(T_0,T_\text{m})\right|^2 
\nonumber\\
& +2 \text{Re}\Big({I_{\bm n}^\text{c}}^*(T_\text{m},T)\Big[e^{-\ii\beta\omega_{\bm n}}I_{\bm 
n}^\text{q}(T_0,T_\text{m})-I_{\bm n}^\text{c}(T_0,T_\text{m})\Big]\Big)\bigg].
\end{align}

The difference of the detector's particle counting in both scenarios 
$\Delta P_\text{e} (T_0,T)$ will be considerable even for 
$T\gg l^3/\lplanck^2$, that is, if we look at the detector 
nowadays. 
Nevertheless, 
{we would probably not be able to resolve times as small as $l^3/\lplanck^2$ 
(roughly, the  Planck scale).} Instead, any observations we may make on particle 
detectors will be averaged in time over many Planck times,
\begin{equation}\label{ave}
\left\langle   P_\text{e}(T_0,T) \right\rangle_{\raro{T}}=\frac{1}{\raro{T} }\int^T_{T-\raro{T}} P_\text{e}(T_0,T')\,\text{d}T',
\end{equation}
where $\raro{T}\gg l^3/\lplanck^2$is the time resolution with which we can probe the 
detector. This 
will partially erase the observable difference between the response of the 
detector in the two regimes. Moreover, in order to remove any 
possible spurious effects coming from the big differences in the  
scales of the problem, we will consider a particle detector with an energy gap 
$\Omega\ll\lplanck^2/l^3$, which additionally 
conspires against the hypothesized visibility of the effect.  Remarkably, and 
contrary to intuition, the difference between the long time averaged response of 
sub-Planckian detectors in the GR and LQC scenarios remains non-negligible even 
under these coarse-graining conditions.

\begin{figure}
\begin{center}
\includegraphics[width=0.44\textwidth]{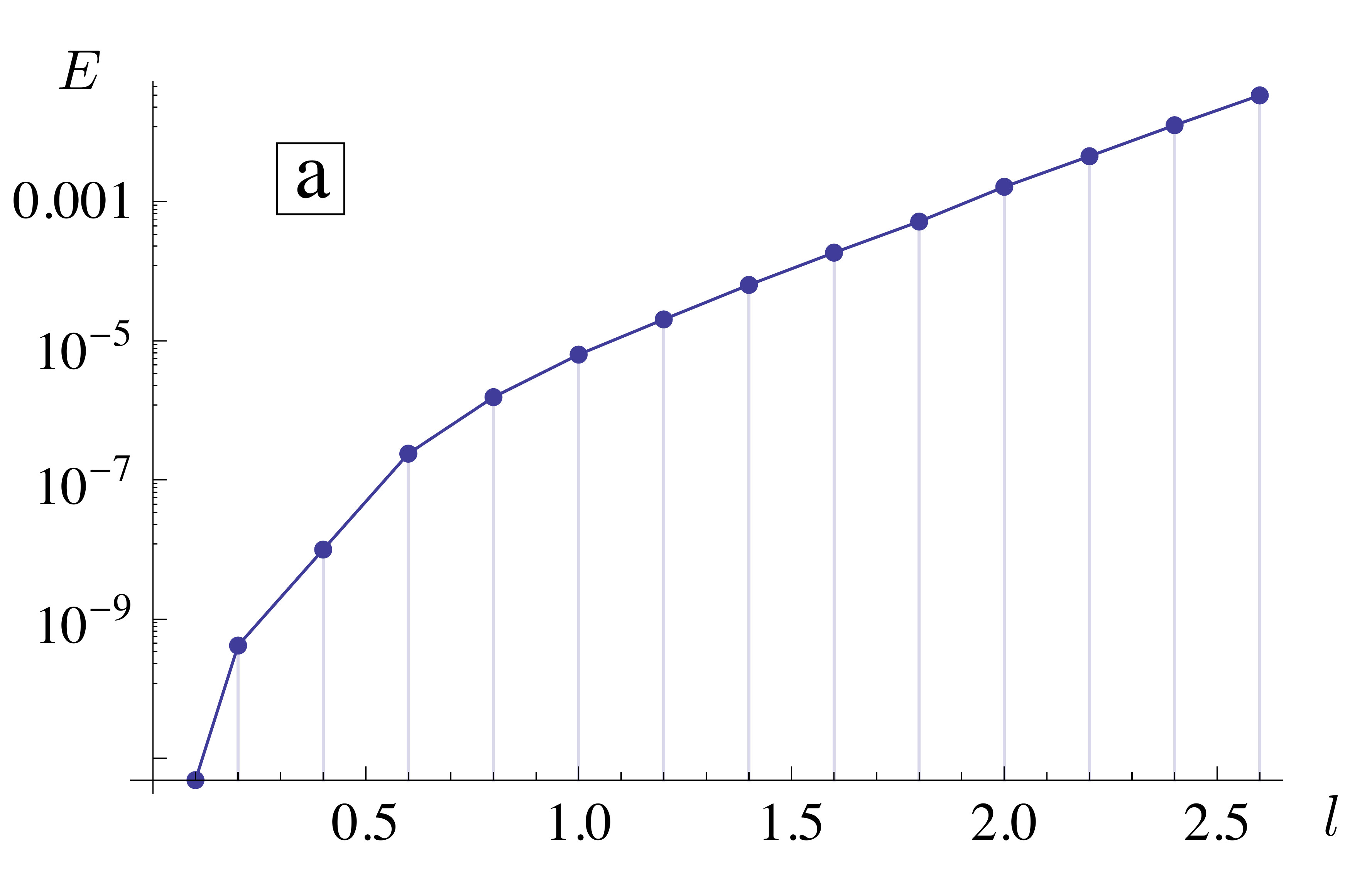}
\includegraphics[width=.44\textwidth]{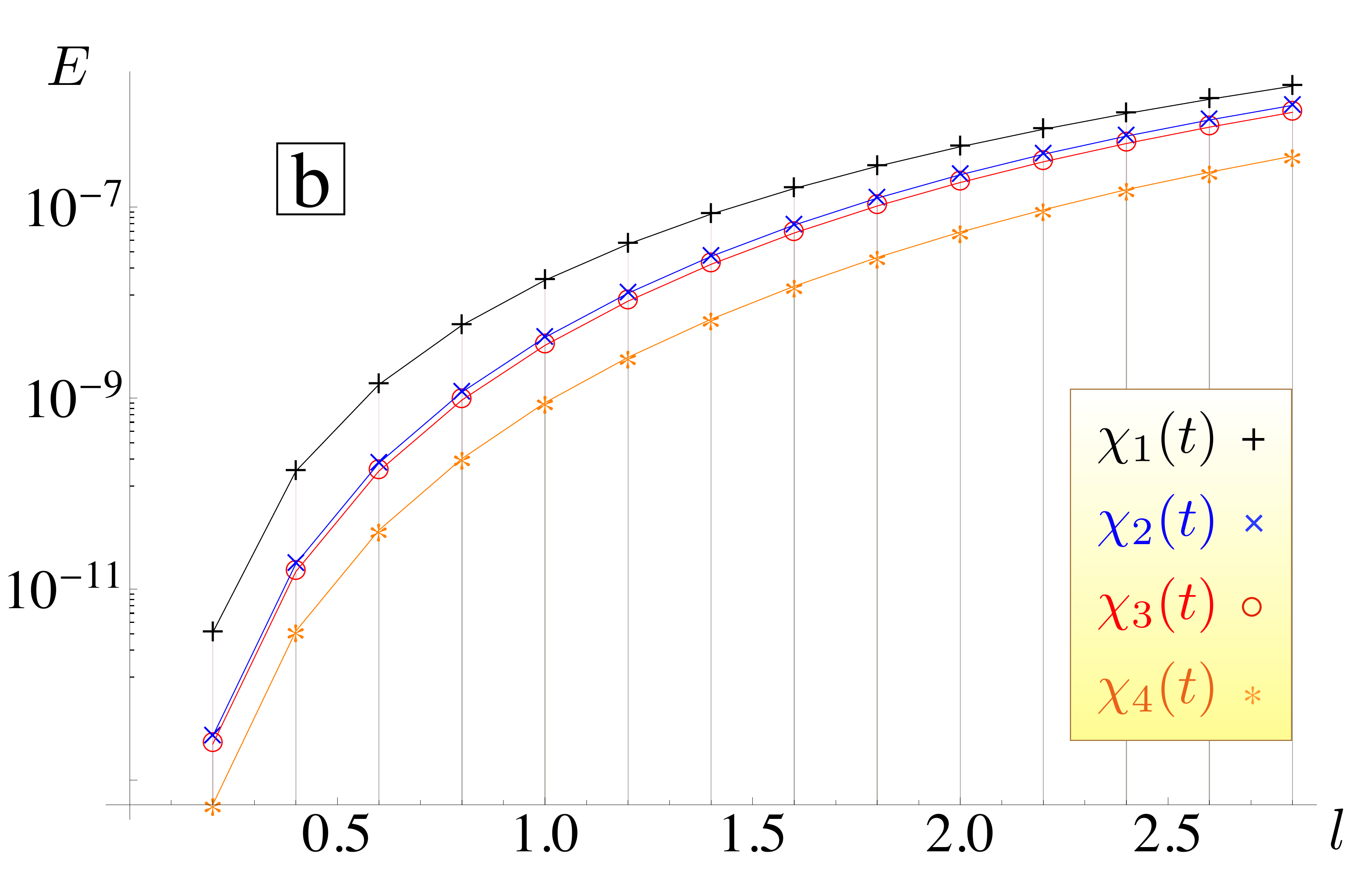}
\caption{[\textbf{(a)}] Logarithmic plot of the relative difference of the averaged probabilities  $E$ (with a regularized denominator) for the sudden switching $\chi_1(t)$
as a function of the parameter $l$, for $\Omega\ll \lplanck^2/l^3$ and $\pi_\varphi=1000$.  The detector is switched on at $T_0=0.01$ (some early time after the bounce). [\textbf{(b)}] Logarithmic plot of the difference estimator $E$ 
as a function of the parameter $l$ for the different switching functions, for a switching time scale $\delta=10$ in 1+1D. The behavior is identical in all four cases.
All dimensionful quantities are expressed in 
Planck 
units.}
\label{fig:size}
\end{center}
\end{figure}

Let us study how sensitive  the response of the detector is to { the LQC quantum parameter $l$  
that characterizes the size of the quanta of volume}. With this aim, let us consider a simple estimator: the mean of the relative difference between probabilities of excitation averaged over a long interval in the late time regime  $\Delta T=T-T_{\text{late}}$, with $\Delta T,\;T_\text{late}\gg l^3/\lplanck^2$:
\begin{align}\label{estimator} 
E=\left\langle\frac{\left\langle  \Delta P_\text{e}(T_0,T) 
\right\rangle_{\raro{T}}}{\left\langle  P^\text{c}_\text{e}(T_0,T)
\right\rangle_{\raro{T}}}\right\rangle_{\Delta T}.\end{align}
This estimator is well defined for all the switching functions above except for the sudden switching $\chi_1(t)$ [as it can be easily checked in formula (3.16) of \cite{Louko:2007mu}]. For this case, it is enough to replace the denominator by $P^\text{c}_\text{e}(T_0,T)_\text{reg} $ which is a UV regularized probability, defined by cutting up the sum in Eq. \eqref{prob} to certain large frequency. This regularization is nonetheless qualitatively irrelevant: We could have used as the regularized denominator in Eq. \eqref{estimator}  the difference with  the probability of excitation of an identically switched detector in Minkowski space-time, obtaining identical results.  The UV cutoff regularization at the Planck scale provides a smaller value for the estimator so, for the sake of being as conservative as possible, we employ that one in Fig. \ref{fig:size}(a). Note that, in a hypothetical experiment, one would measure the absolute difference between the experimentally observed response of the detector and a prediction for the very same detector based on a theoretical cosmological model. We make $E$ a relative difference estimator, however, to get an idea of the magnitude of the differences in the two models. It is easy to check that given that the vacuum excitation is always a bounded oscillatory function, the absolute difference behaves in a very similar way as the estimator defined above.

We will study $E$ as a function of the parameter $l$. This estimator tells us the difference in 
magnitude between the number of clicks of a detector in a LQC background and a 
detector in the classical background.  As shown in Fig. \ref{fig:size}(a), the variation of the response of the detector (the 
intensity of Gibbons-Hawking-type quantum fluctuations) grows exponentially with 
the size of the quantum of volume. This in turn means that the size of the 
quantum cannot be much beyond the Planck scale or the effects would be too large 
nowadays. Although this is a toy model, it captures the essence of a key 
phenomenon:  Quantum field fluctuations are extremely sensitive to the physics of 
the early universe, and the effect survives all the way to the current era. What is 
more important, this exponential dependence on the size of the volume quanta suggests that cosmological observations could put stringent upper bounds to  the quantum scale $l$ in LQC or, equivalently, to the time scale $T_\text{m}$ when the 
quantum effects become  negligible. The exponential growth of the fluctuations 
with $l$ narrows its possible values to a range very close to the Planck scale. 

A very computationally demanding numerical calculation reveals that the same behavior observed for the sudden switching regime is also present in the case of the other smoother switchings $\chi_2(t),\chi_3(t),\chi_4(t)$. For computational convenience and to see how the dimension of space-time affects our result, we show in  Fig. \ref{fig:size}b the estimator $E$ in a 1+1D scenario for the different switchings. Note that  in 1+1D the signal is overall smaller than in the 3+1D. This indicates that the higher dimension amplifies the effect. The main result--exponential dependence of $E(l)$--remains for all four switching functions. In conclusion, the difference between the detector's averaged response in the classical and the quantum scenario depends exponentially on the scale of the quantum of volume and this exponential trend does not depend on the time scale of the detector's activation $\delta$ (smoothness of the switching) or the nature of  the switching function.

\section{Conclusions} 

We found that the Gibbons-Hawking effect is extremely 
sensitive to the physics of the very early universe. A large difference appears between 
an effective model derived from LQC and the classical analog within GR, even if 
after some small time $T_\text{m}$  both dynamics are indistinguishable.  We could, in 
principle, think that the difference between probabilities might decay as 
the time $T$ during which the detector is switched on increases, since the 
larger $T$ is the longer both dynamics coincide. As we showed here, this is not the case
and, most remarkably, the difference survives in time and is extremely 
sensitive to the specifics of the quantum theory, even under very conservative assumptions on the way that the fluctuations are detected. For instance, this may well allow for the derivations 
of stringent upper bounds on the quantum of volume that LQC features.

Although we use the detector just as a witness of quantum fluctuations and the Gibbons-Hawking effect, one can think that generic nonconformal fields  (massive fields, for instance) that couple to the test scalar field will undergo similar effects, therefore impacting inflation and the CMB. Actually, the analysis of inflation within LQC, and how quantum effects might affect the power spectrum of primordial fluctuations has been recently considered \cite{FernandezMendez:2012vi,Fernandez-Mendez:2013jqa,Agullo:2012sh,Ashtekar:2013xka}. Our study gives a different perspective on how to detect signatures of quantum gravity imprinted in the early universe dynamics, and moreover serves to give an upper bound to the LQC quantum scale $l$.

This model can be generalized to {the case of an infinite-level harmonic oscillator detector} whose response at leading order in perturbation theory is the same as that of the two level system \cite{UdWGauss, Fuenetesevolution}. Additionally, the choice of an initial vacuum other than the conformal vacuum used for our particular calculations would quantitatively change the results but one can see that this will not change the main result reported in this paper. Another extension of this work is the analysis of how signatures of quantum {gravity} can be enhanced by studying  the quantum correlations acquired between two different detectors (e.g. exploring the entanglement-harvesting  phenomenon, 
{see e.g.} \cite{Reznik2005,PastFutPRL}).  Additionally, this work sets the basic tools to explore the possibility of transmission of quantum information from the prebounce shrinking universe to the expanding era in a LQC scenario, which will be studied elsewhere.

{\it Acknowledgements}. We are indebted to J. Louko for his extremely helpful comments and sharp insight into the Unruh-DeWitt detector model. L.J.G. and M.M.-B. were partly funded by the  MINECO project FIS2011-30145-C03-02. L.J.G. was also funded by Consolider-Ingenio 2010:  CPAN (CSD2007-00042). He also thanks the hospitality of the University of Waterloo and Perimeter Institute (PI) during his visit in 2012. E.M.-M. was supported by the Banting fellowship programme. Research at PI is supported by the Gov. of
Canada through Industry Canada and by the Province of Ontario through the Ministry of Research and Innovation.

\bibliography{biblio}

\end{document}